\documentclass[AMA,Times1COL]{WileyNJDv5} 

\usepackage{todonotes}
\usepackage{longtable}
\usepackage{booktabs}
\usepackage{subcaption}
\usepackage{placeins}

\newcommand{\mbf}{\mathbf}
\newcommand{\esp}[1]{\mathbb{E}\left\{ #1 \right\}}
\newcommand{\itu}[1]{Rec. ITU-R P.#1}

\articletype{Preprint}%

\received{N/A}
\revised{N/A}
\accepted{N/A}
\journal{Preprint}
\volume{N/A}
\copyyear{N/A}
\startpage{1}

\begin{document}

\title{Improving the estimation of attenuation in Q/V Band systems with a Kalman Based Scintillation Filter}

\author[1]{Justin Cano}
\author[1]{Julien Queyrel}
\author[1]{Laurent Castanet}
\author[2]{Michel Bousquet}

\authormark{CANO \textsc{et al.}}
\titlemark{Improving the estimation of attenuation in Q/V Band systems with a Kalman Based Scintillation Filter}

\address[1]{\orgdiv{DEMR, ONERA}, \orgname{Universit\'e de Toulouse}, \orgaddress{31000 Toulouse, \country{France}}}
\address[2]{ \orgname{MIDIVAL}, \orgaddress{\state{Occitanie}, \country{France}}}

\corres{Corresponding author : Justin CANO, \\	DEMR/ONERA \\ 2 Av. Marc Pellegrin, \\ 31000 Toulouse, France. \email{jcano@onera.fr}}



\abstract[Abstract]{
	
	This paper presents the design and implementation of a Scintillation
	Filter by Kalman-colored algorithm (SciFi), which is used to remove
	tropospheric scintillation in Q/V band total attenuation time series.
	In contrast to the classical methods using low-pass filters, the SciFi
	algorithm allows to estimate both the attenuation, its slope and a
	confidence interval. Moreover, the linear observer structure of the
	Kalman filter allows real-time operation. Therefore, the states
	and uncertainties estimated by SciFi can be used as input for Fade
	Mitigation Techniques (FMT) such as Adaptive Coding and Modulation (ACM)
	or Site Diversity (SD). In this article, a method to tune the
	estimator based on ITU-R Recommendations is proposed. Finally,
	some time-series and statistical results of filtering on Alphasat experimental data are discussed.}


\jnlcitation{\cname{%
\author{Cano J},
\author{Queyrel J},
\author{Castanet L}, and
\author{Bousquet M}}.
\ctitle{Improving the estimation of attenuation in Q/V Band systems with a Kalman Based Scintillation Filter} \cjournal{} \cvol{}
}

\maketitle

\renewcommand\thefootnote{}
\footnotetext{\textbf{Abbreviations:} 
	FMT, Fade Mitigation Techniques; ACM, Adaptive Coding and Modulation; SGD, Smart Gateway Diverity; PSD, Power Spectral Density; KF, Kalman Filter.}

\renewcommand\thefootnote{\fnsymbol{footnote}}
\setcounter{footnote}{1}

\section{Context and Problem Statement}
\label{problem-statement}


Satellite communication systems are 
aiming at delivering higher and higher data rates while
the spectrum becomes more and more congested. To address these two antagonist issues, system designers tend to increase the carrier frequencies to broader and less allocated bandwidth such as Ka, Q/V or W bands \cite{maral_satellite_2020}. However, at such
frequencies above $20~\mathrm{GHz}$, the tropospheric attenuation, mainly due to rain, can far exceeds $20~\mathrm{dB}$, which can lead to service outages. In order to provide a maximal availability of such communication systems, Fade Mitigation
Techniques (FMT) are required to
maintain the system availability during critical rain events \cite{panagopoulos_satellite_2004}. 

However, these techniques require reliable real-time attenuation estimates in order to be effective and to prevent link interruptions. For instance, Adaptive Coding and Modulation (ACM), needs reliable real-time short-term predictions of the channel to trigger an efficient Modulation and Coding (ModCod) scheme. Indeed, increasing the spectral efficiency, of the ModCod decreases its robustness with respect to the channel conditions \cite{bischl_adaptive_2010}. In other terms, a slight raise of the attenuation within some tens of milliseconds can lead to wrong decisions when decoding the incoming signal. Finally, FMTs that involves the comprehensive network, such as the Smart Gateway Diversity (SGD) necessitate to produce mid-term estimates (typically few minutes) in order to anticipate site diversity decisions. In fact, a switch between two gateways necessarily involves terrestrial backbone network operations introducing this delay \cite{ventouras_assessment_2021}.   

Estimating the fade slope is relevant for predicting short-term attenuation
which can be required as FMT input, or to calculate statistics on attenuation dynamics for designing resilient communication systems. 
However, the estimation of the excess attenuation slope, which is mainly driven by rain events, is hampered by the
high-frequency tropospheric scintillation, which introduces a significant noise level in the attenuation time series \cite{comisso_tropospheric_2023}. In order to minimize
erroneous FMT triggering, this phenomenon needs to be filtered out of the attenuation time series and before the attenuation slope is calculated. More directly, this treatment is required to extract the various contributions (gaseous, rain, cloud \textit{etc}...) of the attenuation.

In this paper we address the problem of estimating the excess attenuation and its derivative in real time while acquiring scintillation-noisy measurements. Instead of filtering the scintillation and then estimating the attenuation slope, we propose to solve the problem simultaneously, while providing numerical uncertainties that can be used as additional information for FMT triggering.  Here we constrain the problem by considering only the attenuation time series as input to our estimator, \textit{i.e.}, without any extrinsic measurements such as radiometers or weather radars. This signal is extracted from link budgets between a beacon payload and a receiver at a given location, and must be implemented in real time with simple parameter tuning. The overall objective of this work is the development of a low cost operational algorithm that facilitates the development of new FMTs for next generation satellite communications systems.

\section{Related work}
\label{sec:related_contribution}

In order to design an operative estimator, we require to have a modeling of the scintillation dynamics. First, let us focus on the models proposed by the International Telecommunication Union -- Radiocommunication Sector
(ITU-R), in particular the \itu{618-13} \cite{international_telecomunications_union_propagation_2017} for the statistical modeling, or \itu{1853-2} for the time-series \cite{international_telecomunications_union_time_2019}, which specifies guidelines for the synthesis of total
tropospheric attenuation, including scintillation. In the later document, the scintillation is modeled as a colored (pink) noise,
assuming a low-pass profile with a cutoff frequency of
$f_{c} = 0.1~\mathrm{Hz}$ and an attenuation slope of $-80/3~\mathrm{dB/dec}$.

This frequency profile is explained by the Von Karman-Kolmogorov
Spectrum (VKKS) of refractivity, which provides turbulence strength and
which is detailed in the work of Fabbro \textit{et al.} \cite{fabbro_scintillation_2013}. In particular, it admits a
$({{f/f}_{c})}^{8/3}$ asymptote as the frequency $f$ tends to infinity, where the
cutoff (also referred as ``corner'') frequency is equals to
$f_{c} = 1.43v_{y}\sqrt{2\pi\lambda R}\text{\ .}$ This expression is
parameterized by the wavelength $\lambda$, the wind velocity across
propagation direction $v_{y}$, and the path length in turbulence
$R$. Since such parameters are difficult to measure, the ITU-R recommends in \itu{1853-2} \cite{international_telecomunications_union_time_2019}
a median value of $0.1~\mathrm{Hz}$ for $f_{c}$.

\begin{figure}[h]
	\centering
	\includegraphics[width=0.6\linewidth]{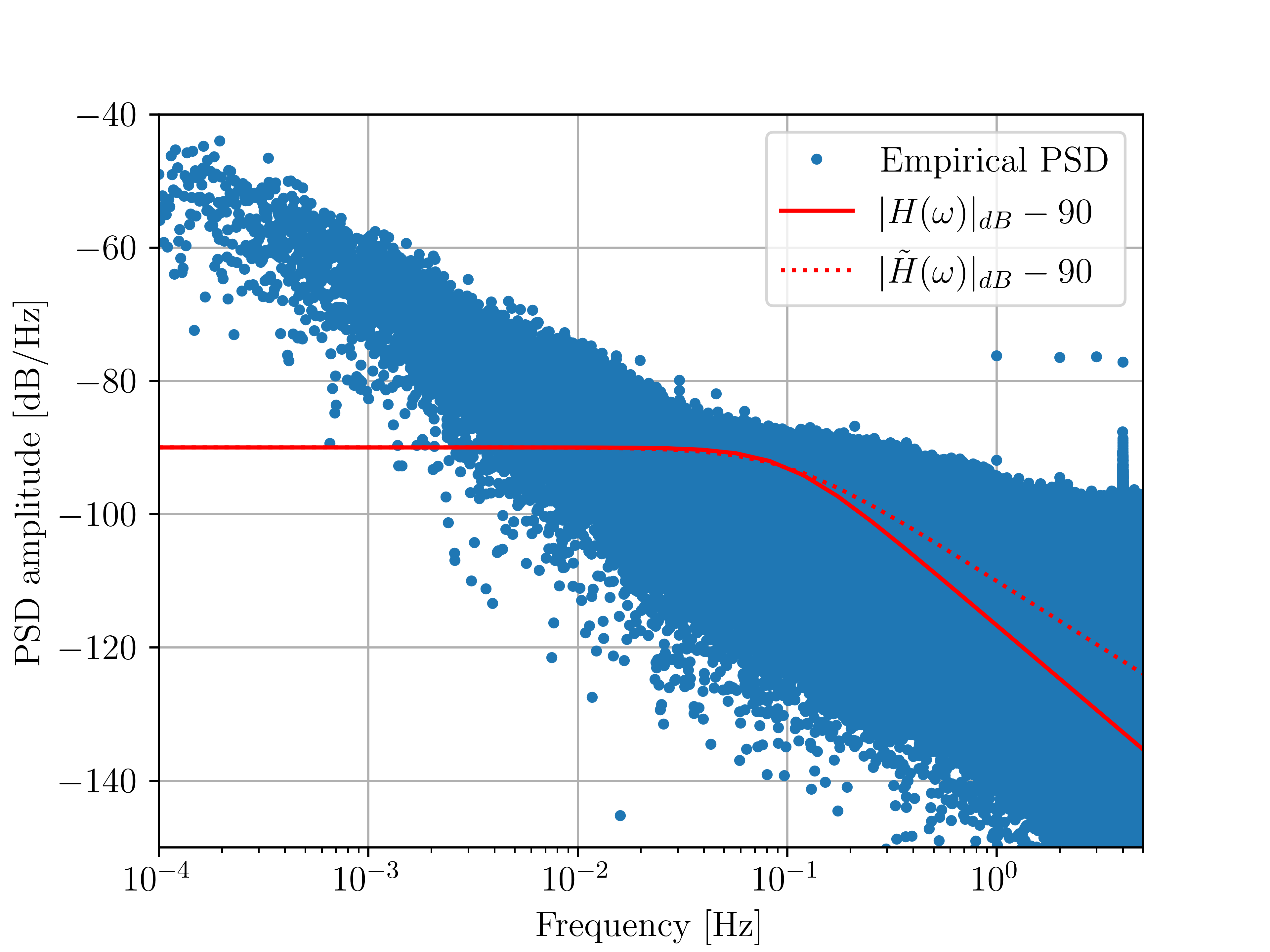}
	\caption{Fourier spectrum of a one-week attenuation time series acquired in Toulouse (France) at Q band ($39.4~\mathrm{GHz}$).}
	\label{fig:spectrum}
\end{figure}

This characteristic spectral profile can be seen in several studies \cite{kelmendi_tropospheric_2022,chakraborty_tropospheric_2021,adhikari_studies_2011,fabbro_scintillation_2013,adhikari_studies_2011}, and is shown in 
Fig. \ref{fig:spectrum}. This plot is the result of Fourier transforming one week of attenuation data from the \emph{Aldo
	Paraboni} campaign (Q band in Toulouse), calculated using the
FFT algorithm. In this figure we can clearly see the
decreasing asymptotic profile predicted by the VKKS. Note that this
includes the total attenuation, which is not exclusively scintillation.
The paper of Kelmendi \textit{et al.} \cite{kelmendi_tropospheric_2022} shows a similar spectrum with a low-pass asymptote of $-80/3
~\mathrm{dB/dec}$, which is consistent with the VKKS, where different frequencies (Q/V and Ka bands) are studied. 

In the literature, the tropospheric scintillation is generally considered to be high frequency noise, which is filtered by low-pass methods, as suggested
in several papers \cite{kelmendi_tropospheric_2022,chakraborty_tropospheric_2021,comisso_tropospheric_2023}. For example, in the work of Charkraborty \textit{et al.} \cite{chakraborty_tropospheric_2021} a high-order Buterworth filter is used to remove tropospheric scintillation from tropical attenuation data measured from GSAT-14 Ka-band beacons. Namely, the authors implemented a tenth-order
Butterworth filter with a cutoff frequency of $0.025~\mathrm{Hz}$, to remove efficiently high frequency noise. 
Alternatively, the approach of Comisso \textit{et al.} presented in \cite{comisso_tropospheric_2023} used
the same scheme, but with a lower-order Butterworth filter for data obtained from temperate regions. 

Overall, while these low-pass filtering approaches are effective in averaging scintillation fluctuations, they introduce a delay in attenuation estimation. This delay, in turn, can cause inaccuracies in FMT control algorithms, adversely affecting the quality of service. Additionally, the low-pass filtering aims at removing the high-frequency components of the signal, potentially eliminating valuable information that could enhance the modeling of the phenomena.

As an alternative to Fourier analysis, the wavelet transform offers a viable method for filtering scintillation. Maitra \textit{et al.} \cite{maitra_scintillations_2008} successfully applied this technique to separate rain and tropospheric scintillation attenuation at Ku-band. Similarly, the wavelet transform has been utilized in satellite navigation systems (GNSS) to identify ionospheric scintillation disturbances, as demonstrated in the paper of Materassi and Mitchell \cite{materassi_wavelet_2007}.
While the wavelet approach is more accurate, it still introduces a delay, making real-time implementation challenging. Moreover, the wavelet transform is more computationally complex than the FFT and does not provide explicit bounds on uncertainties nor fade slope estimates.

Finally, let us review the fade slope estimation techniques. The \itu{1623-1} \cite{itu-r_recommendation_2005}, which is mainly based on the work of Van de Kamp \cite{van_de_kamp_statistical_2003}, gives both fade duration and fade slope estimation techniques. Moreover, these works yield statistical modeling of the fade slope by providing methods to compute the Complementary Cumulative Density Function (CCDF) of the slope conditioned to the attenuation level. 
To estimate the slope, it is assumed that a low-pass filter is used before performing a classical first-order differentiation scheme (also called Euler method). This methodology is used to estimate the fade slope from experimental campaigns in different studies at different locations \cite{boulanger_small_2018,boulanger_four_2015,van_de_kamp_statistical_2003}. 
However, the Euler method, even with low-pass filtering, does not integrate the colored scintillation noise dynamics and the attenuation slopes are averaged over relatively long time periods (several tens of seconds), which introduces lags in the estimates and thus does not allow time-responsive FMT triggering. Furthermore, both attenuation and its derivative are required to compute conditional CCDF of the slope knowing the attenuation level as \itu{1623-1} suggests. The development of an algorithm that estimates the two time-series simultaneously while removing noise would be beneficial to improve statistical modeling of the attenuation slope.
 
\section{Contribution} 

In contrast to the above-mentioned frequency-based algorithms to remove the tropospheric scintillation in post-processing, we propose a real-time approach based on the stochastic
state-space filtering. This kind of method facilitates the simultaneous estimation of the attenuation and its slope while acquiring noisy time series. To do so, we leverage Kalman Filter (KF) framework which is very popular in signal processing to solve such optimization problem in real-time, thanks to their moderate computational cost \cite{chui_kalman_2017,kovvali_introduction_2022}. This method is applied assuming a simple dynamic model observed with noisy measurements, including a modeling of the noise. Therefore, we propose in this article the Scintillation Filter algorithm (SciFi) that:

\begin{itemize}
	\item
	estimates in real-time both the attenuation and its slope while filtering
	scintillation noise;
	\item models the scintillation dynamics as a colored noise;
	\item
	provides estimation uncertainties leveraging KF method;
	\item
	allows to output short-term prediction (and its subsequent
	uncertainty) since the KF allows to compute
	prediction at all time;
	\item allows to compute attenuation and slope statistics at a moderate computational cost; 
\end{itemize}

This paper is organized as follows: Section \ref{sec:kalman_filtering} gives an introduction to Kalman filtering and generalizes its formulation for estimation problems under colored observation noise. Then, Section \ref{sec:scifi} formulates the SciFi algorithm, which provides a physical modeling of the observation, dynamics, and noise tuning based on the ITU-R recommendations that are incorporated into the colored KF framework. Finally, in Section \ref{sec:valid}, we apply SciFi to measured data from an experimental campaign to compute filtered attenuation time series of fade slopes, and their conditional CCDF compared against low-pass/differentiation method.

\subsubsection*{Notation}\label{notation}

In this paper we denote $\hat{x}$ (with a hat) the estimate of a variable
$x$. Vectors are noted with \textbf{bold} font, and $\mbf x^\top$ denotes the vector $\mbf x$ transposed. The time
variable $t$ is written between parenthesis and time indexes are at
subscript, \emph{e.g.}, $x_{k} = x(t_{k})$. We define the time-step
$h_{k}:= t_{k} - t_{k - 1\ }$ that could be modified over time. We also
introduce the notation ``$:=$'' to state that an expression is assigned
``by definition'' or assuming a hypothesis. 
The $n$-th order
identity matrix is denoted as $\mathbf{I}_{n}$. We denote the
time derivatives $\frac{\partial\mathbf{x}}{\partial t} = \dot{x}$
and, respectively
$\ddot{\mathbf{x}} = \frac{\partial^{2}\mathbf{x}}{\partial t^{2}}$
for some continuous time variable $\mathbf{x} \in \mathbb{R}^{n}$.

We note $\mathbb{E}\{\mbf a\}$ the expectation of a random vector $\mbf a \in \mathbb{R}^n$. 
We denote $\mathrm{\mathrm{cov}}(\mathbf{a},\mathbf{b}) = \esp{(\mbf a-\esp{ \mbf a})(\mbf b-\esp{\mbf b})^\top}$, the covariance matrix of a pair
of two variables $\{\mbf a, \mbf b\} \in \mathbb{R}^n \times \mathbb{R}^m$ and we abusively denote the variance
matrix of $\mathbf{a}$ as
$\mathrm{cov}\left( \mathbf{a} \right): = \mathrm{\mathrm{cov}}(\mathbf{a},\mathbf{a})$. 
The independence of a couple random vectors $\{\mbf a, \mbf b\} \in \mathbb{R}^n \times \mathbb{R}^n$  is denoted as $\mbf a \perp \mbf b$.
A random vector $\mbf r$ is called white if and only if $\esp{\mbf r_k}=\mbf 0$ and $\esp{\mbf r_k \mbf r_l ^\top}=\mbf 0$. A random Gaussian vector $\mbf x$ admitting mean $\mbf m$  and covariance $\mathbf P$ is denoted as $\mbf x \sim \mathcal{N}(\mbf m, \mbf P)$.

\section{Kalman Filtering}
\label{sec:kalman_filtering}

The linear observers are mastered since the 1960s thanks to the research in control engineering, and are intensively used in industry for their moderate computational cost and real-time capacities \cite{kovvali_introduction_2022}. These estimators are fed with time series data and
leverage linear transforms in order to estimate a given vector
$\mathbf{x}_{\text{k\ }} = \left\lbrack \ldots x_{\text{ik}}\ldots \right\rbrack^{\top} \in \ \mathbb{R}^{n}$
of state variables $x_{i,k}$ given an observation
$\mathbf{y}_{\mathbf{k}}  = \mathbf{C} _k\mathbf{x}_{\text{k\ }} \in \ \mathbb{R}^{m}$ which linearly depends on the state vector $\mbf x$. However, the observation are noisy in practice and must be filtered in order to provide reliable estimates. 

In this paper, we focus on the Kalman Filter (KF), which solves an optimization
problem under Gaussian observation and model uncertainties for a moderate computational cost.  This algorithm is extensively used for navigation, synchronization, data smoothing and short-term prediction. The KF addresses the following Mean Square Error (MSE) minimization problem :
\begin{equation}
	\hat{\mbf x}_{k|k} = \underset{\hat{\mbf x}_{k|k}^\mathrm{c} \in \mathbb{R}^n }{\mathrm{argmin}} ~ 
	\esp{ \left| \left| \mathbf{C}_k \hat{\mbf x}_{k|k}^\mathrm{c} - \mathbf{y}_{k} \right| \right|^{2}   },
	\label{eq:minmizationMSE}
\end{equation}
under the following hypotheses :
\begin{enumerate}
	\item the state dynamics are known and linear;
	\item the observation is a white Gaussian centered vector $\mathbf{y}_{k}\mathcal{\sim N(}\mathbf{C}_{k}\mathbf{x}_{k},\mathbf{R}_{k})$ that linearly depends on the state $\mbf x_k$;
	\item there exist a prior estimate $\ {\widehat{\mathbf{x}}}_{k-1|k - 1}\mathcal{\sim N(}\mathbf{x}_{k-1},\mathbf{P}_{k-1|k - 1})\ $.
\end{enumerate}

The KF computes the state estimate ${\widehat{\mathbf{x}}}_{k|k}$ recursively and also its covariance matrix $\mathbf{P}_{k|k}$. The recursive character of the algorithm makes it computationally efficient allowing it to be used in real time. The use of the estimate covariance matrix $\mathbf{P}_{k|k}$ enables confidence interval computation, which are two critical arguments in the choice of this mathematical framework for our problem. However, the KF has to be adapted when the observation vector is not noised by white Gaussian noise. Section \ref{ss:kalcolor}, presenting the KF-colored algorithm, assesses specifically this issue.

\label{general-principle}

\subsection{Standard Kalman Filter for white noises}
\label{ss:stdkf}

\begin{figure}
	\centering
	\includegraphics[width=\linewidth]{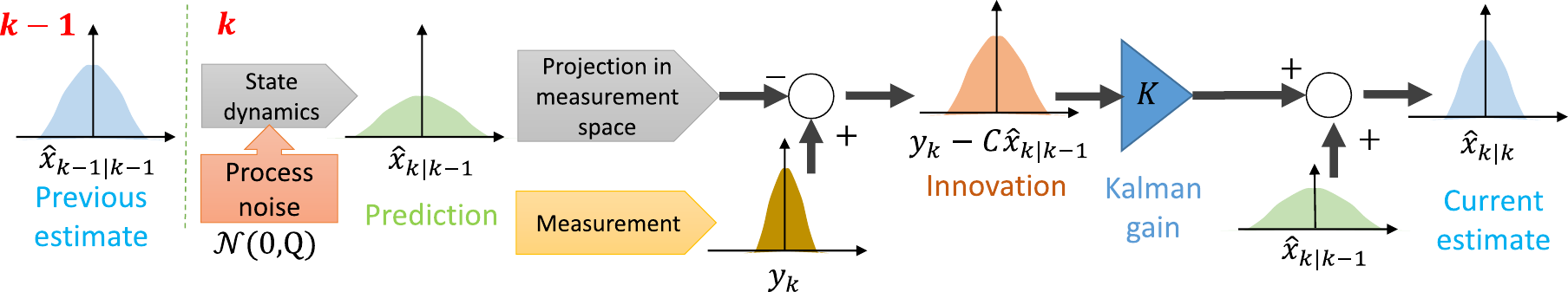}
	\caption{General principle of the Kalman filter.}
	\label{fig:pple_kf}
\end{figure}

%
	 In that case, we assume that the systems dynamic is driven by a recursive equation
		\begin{equation}
			\mathbf{x}_{k} = \mathbf{A}_{k}\mathbf{x}_{k - 1} + \mathbf{w}_{k},
			\label{kf:model}
		\end{equation}
		where $\mathbf{w}_{k} \in \mathbb{R}^{n}$ is the \textit{process noise}, assumed white Gaussian centered such that 
		$\mathbf{w}_{k}\mathcal{\sim N}\left( \mathbf{0},\mathbf{Q}_{k} \right) $
		and $\mathbf{A}_{k} \in \mathbb{R}^{n \times n}$ the dynamics matrix.
		
		Respectively, we consider a linear observation
		\begin{equation}
			\mathbf{y}_{k} = \mathbf{C}_{k}\mathbf{x}_{k} + \boldsymbol{\nu}_{k},
			\label{kf:observation}
		\end{equation}
		with the observation matrix $\mathbf{C}_{k}$
		$\in \mathbb{R}^{n \times m}$ and the Gaussian \textit{measurement noise} 
		$\boldsymbol{\nu}_{k} \in \mathbb{R}^{m}$
		assuming $\boldsymbol{\nu}_{k}\mathcal{\sim N(}\mathbf{0},\mathbf{R}_{k})$.
		
		Note that $\mathbf{R}_{k}$ is ``instantaneous'' since it only depends
		on the measurement quality while $\mathbf{Q}_{k}$ grows with the time
		step $h_k$. Indeed, the uncertainty increases with the
		last estimation time, and $\mathbf{Q}_{k}$ is the covariance matrix
		related to the prior uncertainty.
		\begin{boxwithhead}
			{
			KF filter algorithm	
			}
			{
		The distribution of the initial estimate is assumed Gaussian such that
		${\widehat{\mathbf{x}}}_{0|0}\mathcal{\sim N}\left( \mathbf{x}_{0},\mathbf{Q}_{0} \right)$\emph{,}
		\emph{i.e.,} that the initial mean $\mathbf{x}_{\mathbf{0}}$ and
		covariance $\mathbf{Q}_{0}$ are known. In practice, $\mathbf{x}_{0}$
		is supposed to be equal to the empirical mean of the variables to track
		and that $\mathbf{Q}_{0} = {\kappa\mathbf{Q}}_{k}$ where
		$\kappa \gg 1$ and a standard $\mathbf{Q}_{k}$ given a nominal time
		step $h_{k}$.
		
		The Kalman Filter is a recursive algorithm that can be divided into
		three steps, assuming $k \geq 1$:
		
		\begin{enumerate}
			\item
				The \textbf{prediction step}, that builds the prior estimate $\widehat{\mathbf{x}}_{k|k - 1}$ and its covariance $\mbf P_{k|k-1}$
				\begin{equation}
					\begin{cases}
						{\widehat{\mathbf{x}}}_{k|k - 1}&=\mathbf{A}_{k}{\widehat{\mathbf{x}}}_{k - 1\mathbf{|}k - 1}, \\ 
						\mathbf{P}_{k|k - 1}&:= \mathrm{cov}(\mbf x_{k|k-1}) =  \mathbf{A}_{k}\mathbf{P}_{k - 1\mathbf{|}k - 1}\mathbf{A}_{k}^{\mathbf{\top}}\mathbf{+}\mathbf{Q}_{k};
					\end{cases}
					\label{eq:kf_pred}
				\end{equation} 
		
			\item		The \textbf{innovation step} : that compares the prediction to the incoming
				measurement, which consist in the computation of the innovation $\mathbf{i}_{k|k - 1}$ (and its covariance) as follows 
				\begin{equation}
					\begin{cases}
						\mathbf{i}_{k} &= \mathbf{y}_{k}\mathbf{-}\mathbf{C}_{k}{\widehat{\mathbf{x}}}_{k|k - 1}, \\
						\mathbf{S}_{k} &:=
						\mathrm{cov}(\mbf i_k)
						=
						\mathbf{R}_{k}\mathbf{+}\mathbf{C}_{k}\mathbf{P}_{k|k - 1}\mathbf{C}_{k}^{\mathbf{\top}};
					\end{cases}
				\end{equation}
		\item Finally, \textbf{the update step} consists in merging the information obtained thanks to the measurements (innovation) and prior estimates (prediction) 
		\begin{equation}
			\begin{cases}
				{\widehat{\mathbf{x}}}_{k|k}&={\widehat{\mathbf{x}}}_{k|k - 1}\mathbf{+}\mathbf{K}_{k}\mathbf{i}_{k}, \\ 
				
				\mathbf{P}_{k|k} &:= \mathrm{cov}(\hat{\mbf x}_{k|k}) = {\mathbf{(}\mathbf{I}_{n}\mathbf{- K}}_{k}\mathbf{C}_{k}\mathbf{)}\mathbf{P}_{k|k - 1}.
			\end{cases}
		\end{equation}
		This step leverages the \textit{Kalman gain} 
		\begin{equation}
			\mathbf{K}_{k} := \mathbf{P}_{k|k - 1}\mathbf{C}_{k}^{\mathbf{\top}}\mathbf{S}_{k}^{\mathbf{- 1}},
			\label{eq:kalman_gain}
		\end{equation}
		which selects an optimal amount of innovation to be considered in the final
				estimate. The value of $\mathbf{K}_{k}$ is obtained by minimizing the
				trace of $\mathbf{P}_{k|k}$ which is equivalent to minimize the MSE of ${\widehat{\mathbf{x}}}_{k|k}$ as in \eqref{eq:minmizationMSE}
				 under centered distribution assumption.  A proof of optimality can be found in the textbook of Chui and Chen \cite{chui_kalman_2017}. Overall, the KF computes the best compromise between the new information brought by the measurement and the dynamic model enriched by past observations under white Gaussian assumptions. 
	\end{enumerate}
}
\end{boxwithhead}

\subsection{Adapting the KF to colored measurement noises}
\label{ss:kalcolor}

However, the previous equations are not relevant if the observation
noise is not white. In particular, the computation of the Kalman gain in \eqref{eq:kalman_gain} which allows to solve \eqref{eq:minmizationMSE} has been simplified by the fact that the measurement noise was uncorrelated with its past value. In our problem, as highlighted by Fig. \ref{fig:spectrum}, the measurement noise is clearly colored, which makes \eqref{kf:observation} not applicable. Here, we \textit{assume} that we can model the measurement noise by linear dynamics, \textit{i.e.}, a matrix time-recursive equation fed by white noise. In such cases, the algorithm presented in the paper of Chang \cite{chang_kalman_2014}, which is an extension of the KF
that leverages Bryson measurement differentiation \cite{bryson_linear_1965}, can be
used. 

Let assume now that the observation vector $\mathbf{y}_{k}$ is disturbed by a colored noise $\boldsymbol{n}_k$ and that the system driven by the three equations as follows
\begin{subequations}
	\begin{align}
		\mathbf{x}_{k}\mathbf{= \ }\mathbf{A}_{k - 1}\mathbf{x}_{k - 1} + \mathbf{w}_{k - 1}, \label{eq:state_col}\\
		\mathbf{y}_{k}\mathbf{=}\mathbf{C}_{k}\mathbf{x}_{k} + \mathbf{n}_{k}, \label{eq:mes_col} \\
		\mathbf{n}_{k} = \mathbf{F}_{k}\mathbf{n}_{k - 1} + \boldsymbol{\epsilon}_{k - 1} \label{eq:noise_dyna}. 
	\end{align}
\end{subequations}

The equation \eqref{eq:state_col} is unchanged and identical to \eqref{kf:model} where $\mathbf{w}_{k}\mathcal{\sim N}\left( \mathbf{0},\mathbf{Q}_{k} \right)$ is white Gaussian centered. In \eqref{eq:mes_col}, $\mathbf{n}_{k} \in \mathbb{R}^m$ is colored and its dynamics are described in \eqref{eq:noise_dyna}, where $\mbf F_k \in \mathbb{R}^{mm}$ is known and the noise
$\boldsymbol{\epsilon}_{k}\mathcal{\sim N}\left( \mathbf{0},\mathbf{R}_{k} \right)$ is assumed white. Moreover, we assume that $\boldsymbol{\epsilon}_{k}$ and $\mathbf{w}_{k}$ are
independent.  

To tackle the colored character of the noise, we use the Bryson differentiation approach which aims to compensate the noise correlation.  We can compute the following pseudo-observation, thanks to past and present measurements 
\begin{equation}
	\mathbf{z}_{k} := \mathbf{y}_{k} - \mathbf{F}_{k}\mathbf{y}_{k - 1} 
\end{equation}
which can also be rewritten as a function of the previous state leveraging \eqref{eq:state_col}, \eqref{eq:mes_col} and \eqref{eq:noise_dyna}
\begin{equation}
	\mathbf{z}_k = \mathbf{G}_{k}\mathbf{x}_{k - 1} + \mathbf{e}_{k},
	 \label{eq:pseudomeasurement}
\end{equation}
defining the matrix
$\mathbf{G}_{k}: = \mathbf{C}_{k}\mathbf{A}_{k - 1} - \mathbf{F}_{k - 1}\mathbf{C}_{k - 1}$ and the pseudo-noise vector
$\mathbf{e}_{k} = \mathbf{C}_{k}\mathbf{w}_{k - 1} + \boldsymbol{\epsilon}_{k - 1}$, which is a white Gaussian noise vector.
We compute the covariance of $\mathbf{e}_{k}$, while $\boldsymbol \epsilon_{k-1} \perp \mbf w_{k-1}$ as
\begin{equation}
	{\widetilde{\mathbf{R}}}_{k}: = \mathrm{cov}\left( \mathbf{e}_{k} \right) = \mathbf{C}_{k}\mathbf{Q}_{k}\mathbf{C}_{k}^{\top} + \mathbf{R}_{k}
	\label{eq:cov_pseudomeas}
\end{equation}
The pseudo-measurement equation allowed us to define an equivalent system of two equations with known parameterization and white centered random vectors. Therefore, the standard Kalman approach can be implemented.
Let us build the prediction at time $k-1$ and compare it to the pseudo-measurement at time $k$, which is built from data captured at times $k$ and $k-1$.  
First, we can compute the innovation $\mathbf{i}_{k}$, using the previous
estimate ${\widehat{\mathbf{x}}}_{k - 1|k - 1}$ and its covariance
$\mathbf{P}_{k - 1|k - 1}$ as follows
\begin{equation}
	\mathbf{i}_{k} := \mathbf{z}_{k} - {\widehat{\mathbf{z}}}_{k|k - 1} = \ \mathbf{z}_{k} - \mathbf{G}_{k}{\widehat{\mathbf{x}}}_{k - 1|k - 1}.
	\label{eq:innov}
\end{equation}
Respectively, the covariance of the innovation (the pseudo-measurement is independent from the state at $k-1$) is equal to 
\begin{equation}
	\mathbf{S}_{k} := \mathrm{cov}(\mathbf{i}_{k})  = \mathbf{G}_{k}\mathbf{P}_{k - 1|k - 1}\mathbf{G}_{k}^{\top} + {\widetilde{\mathbf{R}}}_{k}.
	\label{eq:innov_cov}
\end{equation}

The cross covariance between the previous estimate and the innovation is
$\mathbf{P}_{xi,k} = \mathbf{P}_{k - 1|k - 1}\mathbf{G}_{k}^{\top}$
, which allows to compute the Kalman gain
$\mathbf{K}_{k} = \mathbf{P}_{xi,k}\mathbf{S}_{k}^{- 1}$ to estimate the state at time $k-1$. The
predicted estimate at time $k - 1$ leveraging both measurements at
$k$ and $k - 1$, the prior estimates and the innovation is built as
follows

\begin{equation}
	\begin{cases}
			{\widehat{\mathbf{x}}}_{k - 1|k} &= {\widehat{\mathbf{x}}}_{k - 1|k - 1} + \mathbf{K}_{k}\mathbf{i}_{k}, \\ 
				\mathbf{P}_{k - 1|k} &:= \mathrm{cov}(\hat{\mbf x}_{k-1|k}) = \mathbf{P}_{k - 1|k - 1} - \mathbf{K}_{k}\mathbf{S}_{k}\mathbf{K}_{k}^{\top}.
				
	\end{cases}
	\label{eq:col_pred}
\end{equation}

The step performed in \eqref{eq:col_pred} allows the incorporation of the noise dynamics to the model.

However, since $\mathbf{w}_{k}$ and $\mathbf{e}_{k}$ are
correlated, the system dynamics cannot be used directly to propagate the estimate at time $k$ (namely, $\mathrm{cov}\left( \mathbf{w}_{k},\mathbf{\epsilon}_{k} \right) = \mathbf{Q}_{k}\mathbf{C}_{k}^{\top} \neq \mbf 0$)
. An algebraic operation that compensates such correlation can be obtained by adding a zero vector obtained from \eqref{eq:pseudomeasurement} to the state equation \eqref{eq:state_col} 
\begin{equation}
	\mathbf{x}_{k} :=  
	\mathbf{A}_{k - 1}\mathbf{x}_{k - 1} + \mathbf{w}_{k - 1} 
	= \mathbf{A}_{k - 1}\mathbf{x}_{k - 1} + \mathbf{w}_{k - 1} + \mathbf{P}_{xi,k}{\widetilde{\mathbf{R}}}_{k}^{- 1} ( \underbrace{\mathbf{z}_{k} - \mathbf{G}\mathbf{x}_{k - 1} - \mathbf{e}_{k}}_{=\mbf 0} )
	= {\widetilde{\mathbf{F}}}_{k}\mathbf{x}_{k - 1} + \mathbf{P}_{xi,k}{\widetilde{\mathbf{R}}}_{k}^{- 1}\mathbf{z}_{k} + \boldsymbol{\xi}_{k},
\end{equation}
${\widetilde{\mathbf{F}}}_{k}: = \mathbf{A}_{k} - \mathbf{P}_{xi,k}{\widetilde{\mathbf{R}}}_{k}^{- 1}\mathbf{G}_{k}$ and
where $\boldsymbol{\xi}_{k} := \mathbf{w}_{k}\mathbf{-}\mathbf{P}_{xi,k}{\widetilde{\mathbf{R}}}_{k}^{- 1}\mathbf{e}_{k}$ is the noise term. 
We then can verify that the previous form of $\mathbf{x}_{k}$ and the pseudo-observation $\mbf z_k$ are not correlated by computing  
\begin{equation}
	\mathrm{cov}(\boldsymbol{\xi}_k,\mbf e_k)= \esp{(\mbf w_k - \mbf P_{xi,k} \tilde{\mbf R}^{-1} \mbf e_k)\mbf e_k^\top}
	=
	\mbf P_{xi,k} - \mbf P_{xi,k} \tilde{\mbf R}^{-1}_k \tilde{\mbf R}_k = \mbf 0, 
\end{equation}
which proves that the decorrelation step has been performed correctly. 
Therefore, the final estimate and its covariance can be obtained by computing  

\begin{equation}
	\begin{cases}
		{\widehat{\mathbf{x}}}_{k|k} &= {\widetilde{\mathbf{F}}}_{k}{\widehat{\mathbf{x}}}_{k - 1|k} + \mathbf{P}_{xi,k}{\widetilde{\mathbf{R}}}_{k}^{- 1}\mathbf{z}_{k}, \\
		\mathbf{P}_{k|k} &= {{\widetilde{\mathbf{F}}}_{k}\mathbf{P}}_{k - 1|k}{\widetilde{\mathbf{F}}}_{k}^{\top} + \mathbf{Q}_{k} + \mathbf{P}_{xi,k}{\widetilde{\mathbf{R}}}_{k}^{- 1}\mathbf{P}_{xi,k}^{\mathbf{\top}}. \\
	\end{cases}
	\label{eq:col_propag}
\end{equation}
Given an initial estimate $\hat{\mbf x}_{0|0}$ and its covariance matrix $\mbf P_{0|0}$, the systems \eqref{eq:col_pred} and \eqref{eq:col_propag} are used recursively to estimate $\hat{\mbf x}_{k|k}$ when a new measurement point $\mbf y_k$ is available.

\section{Implementation of the SciFI
		estimator}
\label{sec:scifi}

Here, we explain how to use the colored version of the Kalman filter in order to estimate the rain attenuation and its slope under observations noised by tropospheric scintillation.  
To achieve such implementation of the algorithm, named \textit{SciFi} (Scintillation Filter), we need to parameterize Equations \eqref{eq:state_col}, \eqref{eq:mes_col} and \eqref{eq:noise_dyna} presented at the beginning of Section \ref{ss:kalcolor}. In particular, Subsection \ref{ss:attdyn} explains the attenuation dynamics, Subsection \ref{eq:scintillation_dynamics} models the scintillation noise while Subsection \ref{ss:noises} explains how to tune the observation and process noises covariances.   

\subsection{Attenuation dynamics}
\label{ss:attdyn}

We are interested in tracking both the attenuation $a(t)$ and its time
derivative $\dot{a}(t)$, which are therefore considered as the state variables. Then, one can write the continuous-time differential system

\begin{equation}
	\dot{\mathbf{x}} (t) = \mathbf{M}\mathbf{x}_{a} (t) + \mathbf{w} (t), \text{ with } 
	\mbf 
	M
	=
	\begin{bmatrix}
		0 & 1 \\
		0 & 0
	\end{bmatrix}, 
	\mbf w(t) = 
	\begin{bmatrix}
		0 \\ 
		w_{\ddot{a}}(t)
	\end{bmatrix}
	\text{ and }
	\mbf x = 
	\begin{bmatrix}
		a \\ \dot{a} 
	\end{bmatrix}.
	\label{eq:continuous_time_attenuation_dynamics}
\end{equation}

The vector
$\mathbf{w}(t)$
is a continuous-time centered white Gaussian noise vector with an active
component $w_{\ddot{a}}(t)$ on~$\ddot{a}$, which has a Power Spectral Density (PSD) equals to
$s_{\text{ww}}^{2}$. Therefore, the attenuation dynamics \eqref{eq:continuous_time_attenuation_dynamics} describes a Brownian
motion driven by a random Gaussian attenuation ``acceleration'' $w_{\ddot{a}}(t)$, in the absence of extrinsic information. This
basic modeling allows us to tune the level of the variation rate of the
attenuation slope (which remains unknown) thanks to $s_{\text{ww}}$.

In order to implement it into the KF, the differential system \eqref{eq:continuous_time_attenuation_dynamics} can be discretized as follows:
\begin{equation}
	\mathbf{x}_{k} = {\mathbf{A}_{k}\mathbf{x}}_{k - 1} + \mathbf{w}_{a,k}\mathbf{,}
\end{equation}
where
$\mathbf{A}_{k} := \exp{\mathbf{(M}h_{k})}= \mathbf{I}_{2}\mathbf{+}\mathbf{M}h_{k}$
since $\mathbf{M}$ is a one-order nilpotent matrix. The discretization of the process white noise $\mbf w (t)$ requires to compute the covariance of its discrete version $\mbf w_k$, namely 
\begin{equation}
	\mbf Q_k := \mathrm{cov}(\mbf w_k) := \int_{t=0}^{h_k}
	\exp(\mbf Mt) 
	\begin{bmatrix}
		0 & 0 \\
		0 & s_{ww}^2
	\end{bmatrix}
	\exp(\mbf Mt)^\top \mathrm{d}t
	=
	\begin{bmatrix}
		h_k^3/3 & h_k^2/2 \\
		h_k^2/2 & h_k
	\end{bmatrix}
	s_{ww}^2. 
	\label{eq:qmatrix}
\end{equation}


	\subsection{Scintillation noise dynamics}
\label{eq:scintillation_dynamics}

Since the scintillation noise is not a white noise, we have to model its
dynamics in the state-space equations. Namely, the attenuation
observation equation is as follows
\begin{equation}
	y_{k} = a_{k} + n_{k},
	\label{eq:measure}
\end{equation}
where $n_{k}$ is a colored noise. Note that \eqref{eq:measure} immediately yields
$\mathbf{C}_{k}: = \lbrack 1,0\rbrack$, according to the definition of
the state $\mathbf{x}_{k}$.

A colored noise can be seen as the output $n$ of a filter $H$ with
white noise input $\epsilon$. As specified in Section \ref{sec:related_contribution} and \itu{1853-2} \cite{international_telecomunications_union_time_2019},
the slope of such filter is equal to $- 80/3~\mathrm{dB/dec}$ while its
cut-off frequency is $f_{c} = 0.1~\mathrm{Hz}$. Note that $f_c$ should be adjusted, especially at low elevations, \textit{e.g.} see the Ph.D. dissertation of Rytir \cite{phd_rytir}.  However, in order to
simplify the computation of KF iterations and to overestimate high
frequencies components of the noise, we propose to consider a
first-order low-pass filter $\widetilde{H}$. This overestimation
	can be seen in Figure 1, where both the amplitude frequency responses
of $\widetilde{H}$ and $H$ are superimposed. This
	simplification assumption is relevant since other sources of
	high-frequency noises have to be considered (\textit{e.g.}, windowing effects,
	data interpolation, thermal fluctuations).

The discrete (or Z) transfer function of such a filter can be written as
~$\widetilde{H}\left( z \right) = \frac{n}{\epsilon}\left( z \right) = \frac{\text{Kf}_{c}}{1 - e^{- h_{k}f_{c}}z^{- 1}}$
where $z\in \mathbb{ C}$, remembering that $h_{k}$ plays the
role of the sampling period \cite{moudgalya2007digital}. In order to simplify the tuning,
we set $K := f_{c}^{- 1}$. Since $z^{- 1}$ is the delay
	operator in the Z-space, the expression of the transfer function yields to the
	following recursion equation
\begin{equation}
	n_{k} = e^{- h_{k}f_{c}}n_{k - 1} + \epsilon_{k} \text{, where }  \epsilon_{k}\mathcal{\sim \ N}(0,R_{k})
	\label{eq:scintillation_noise}
\end{equation}
the parameterization
of $R_{k}$ is discussed in the next paragraph of this section.
Finally, the equation \eqref{eq:scintillation_noise} gives $F_{k}: = e^{- h_{k}f_{c}}$, which allows to iterate \eqref{eq:col_propag}.

\subsection{Tuning of the noise levels}
\label{ss:noises}

The value of the scintillation variance  $R_k$  is
depending on the location $\mathbf{q}$ (elevation, latitude and
longitude) of the receiving antenna, its efficiency $\eta$, the
operating frequency $f$, and the average radio refractivity wet term
$N_{\text{wet}}$, \textit{e.g.} \emph{see} \itu{618-14} \cite{international_telecomunications_union_propagation_2017} for
details. For the computation of the $N_{\text{wet}}$ term, \itu{453-14} \cite{international_telecomunications_union_radio_2019} can be used for fine computation or at least median values of
that term, extracted from a global map in the Recommendation. On the other hand,
$R_k$ also depends on the attenuation level, resp. \itu{1853-2} \cite{international_telecomunications_union_time_2019}. In
order to have a realistic model for the noise amplitude that is driving
the scintillation, we propose the multiplicative model as follows:
\begin{equation}
	R_{k}: = \sigma_{\mathrm{P.618}}^{2}\left( \mathbf{q},\eta,f,N_{\text{wet}} \right)r({\widehat{a}}_{k - 1|k - 1})
	\label{eq:rmatrix}
\end{equation}

where $\sigma_{P.618}^{2}\left( \mathbf{q} \right)$ is computed using
the methodology presented in \itu{618-14}  \cite{international_telecomunications_union_propagation_2017}, Section 2.4, and we
assume that
$r\left( a \right): = \max (\alpha_{\min},a^{\frac{5}{12}})$
as suggested in \itu{1853-2} \cite{international_telecomunications_union_time_2019}. The parameter $\alpha_{\min}$,
which can be seen as a minimal threshold of scintillation would be tuned
as $\alpha_{\min}: = 2^{5/12}$ to match a $2~\mathrm{dB}$ lower
bound.

Finally, the very last variable to set is the process noise PSD
$s_{\text{ww}}^{2}$ that drives the attenuation dynamics. This
variable is determining the smoothness of the KF, \emph{i.e.,} if this
PSD is increased then the confidence in the model will decrease, thus
scintillation would be less smoothed. We propose to tune it by solving a
least-squares problem during a standard rain event.

\begin{equation}
	\sigma_{\text{ww}}^{*} = \underset{\sigma_{ww > 0}}{\mathrm{argmin}} ~ {\ \sum_{k \in \lbrack 1,E_{k}\rbrack}^{}{\left| {\widehat{a}}_{k|k} - {\overline{y}}_{k} \right|^{2}\ }},
	 \label{eq:q_tuning_ls_optimization}
\end{equation}

assuming that the event occurs between $k$ and $E_{k}$ using a
sliding window mean ${\overline{y}}_{k}$ of twenty points. The problem \eqref{eq:q_tuning_ls_optimization} can be solved using standard Python least-squares solvers, such as those provided in the \texttt{scipy.optimize} package. 

\subsection{Summary of the computation steps}
\begin{algorithm}[h]
	\caption{\enskip SciFi pseudo-code}\label{alg:scifi}
	\begin{algorithmic}
			\State \textbf{Input Data:} $\hat{\mbf x}_{0|0}, \mbf P_{0|0}$
		\State \textbf{Initialization} $a_0:=0$ \textbf{Acquire} $a_1$ and \textbf{Compute} $h_1=t_1$;
		\For {$k>1$} 
		\State \textbf{Acquire} $a_k$ and \textbf{Compute} $h_k=t_k-t_{k-1}$;
		\State \textbf{Update} \textnormal{noise levels} $\mbf Q_k$ and $R_k$ with \eqref{eq:rmatrix} and \eqref{eq:qmatrix}; 
		\State \textbf{Compute} \textnormal{pseudo-measurement} $z_k = a_k - \exp(-h_k f_c)a_{k-1}$ and \textnormal{its covariance} $\bar{R}_k$ with \eqref{eq:cov_pseudomeas};
		\State \textbf{Compute} \textnormal{innovation term} $i_k$ and \textnormal{its covariance} $S_k$ using \eqref{eq:innov} and \eqref{eq:innov_cov};
		\State \textbf{Compute} \textnormal{posterior state estimate} $\hat{\mbf x}_{k-1|k}$ and its \textnormal{covariance} $\mbf P_{k-1|k}$ using \eqref{eq:col_pred};
		\State \textbf{Compute} \textnormal{estimate} $\hat{\mbf x}_{k|k}$ and \textnormal{its covariance} $\mbf P_{k|k}$ using \eqref{eq:col_propag}.
		\EndFor
	\end{algorithmic}
\end{algorithm}

The procedure Algorithm \ref{alg:scifi} summarizes the SciFi processing steps, which leverages previously defined equations. For the initialization, we suppose that the couple $\{\hat{\mbf x}_{0|0}, \mbf P_{0|0}\}$ is known, corresponding to statistical mean and a high covariance value by setting $\mbf P_{0|0}=\mbf Q_{0}$ with $h_0>>1$ in \eqref{eq:qmatrix}, which correspond a covariance estimate not available for a large period of time. This approach is similar to the initial step of the standard KF explained in Subsection \ref{ss:stdkf}. 

	\begin{boxwithhead}
	{
		How to do short term predictions with SciFi? 	
	}
	{
		Assuming that $\hat{\mbf x}_{k|k}$ is given by the execution of \eqref{eq:col_propag} at time $k$, while measuring the last available measurement $y_k$, we aim to compute $\hat{\mbf x}_{k+1|k}$, without new measurement point. In other words,  $\hat{\mbf x}_{k+1|k}$ is the estimate of $\mbf x (t_k+h_k)$ based on the measurement history at $t_k$, since there is no new information we compute the expectation, leveraging \eqref{eq:state_col}, we have
		\begin{equation}
			\hat{\mbf x}_{k+1|k} = \esp{ \mbf A_k \mbf x_k + \mbf w_k | y_k \dots y_1} = \mbf A_k \hat{\mbf x}_{k|k}.
			\label{eq:pred_short_term}
		\end{equation} 
		 The previous computation simply yields the prediction step of the classical KF filter \eqref{eq:kf_pred} that also gives $\mbf P_{k+1|k}$, which is $\hat{\mbf x}_{k+1|k}$ covariance. Indeed, the process noise is assumed white and no further colored measurements are considered in \eqref{eq:pred_short_term}. 
	}
	\end{boxwithhead} 
\section{Filtering Results with experimental data}
\label{sec:valid} 

In this section, we present the results of practical SciFi use-cases on excess attenuation data measured during a multi-year experimental campaign. Namely, in Subsection \ref{ss:used_data}, we describe the dataset used to perform time-series filtering in \ref{ss:real-time} and to compute a statistical analysis over five years of data in \ref{ss:statistic}.   
We implemented SciFi in Python \texttt{3.12} leveraging only the canonical Numpy library for the computation and complementary standard libraries  for the input/output management of the algorithm and the display  (\textit{e.g.}, \texttt{Matplotlib}). 

\subsection{Input data and noise level calibration}
\label{ss:used_data}
In order to validate SciFi implementation, we used attenuation data captured during the \emph{Aldo Paraboni}
experimental campaign. The Aldo Paraboni payload, a reference beacon operating at Q band, is carried by the Inmarsat GEO satellite Alphasat and is part of a European Space Agency experimental program studying tropospheric impairments. 
On the other hand, the ground station proving measurements is located in Toulouse (France) during the
period 2018-2022. 
The input parameters of our SciFi runs are shown in Table
\ref{tab:input}. 
We tuned $\sigma_{\text{ww}}$ by solving the least-squares
problem presented in \eqref{eq:q_tuning_ls_optimization}, for a given rain event present in the dataset.
We recommend using rain events to tune this PSD because the slope value is higher during these events, providing a conservative estimate that allows the attenuation to be correctly tracked. 
The input parameter $N_\mathrm{wet}$ is given by the \itu{453} \cite{international_telecomunications_union_radio_2019}  as specified in the methodology detailed in Section \ref{ss:noises}.

\begin{table}[h]
\centering
\caption{Input parameters for SciFi runs.}
\label{tab:input}
\begin{longtable}[]{@{}llll@{}}
	\toprule
	Variable & Symbol & Value & Unit\tabularnewline
	\midrule
	\endhead
	Elevation angle & $\theta$ & 34.5 & degree\tabularnewline
	Antenna efficiency & $\eta$ & 0.6 & N/A\tabularnewline
	Antenna diameter & $D$ & 1.2 & meter\tabularnewline
	Beacon frequency & $f$ & 39.4 & GHz\tabularnewline
	Median wet term & $N_{\text{wet}}$ & 50 & N/A\tabularnewline
	Process noise PSD & $\sigma_{\text{ww}}$ & ${5.10}^{- 6}$ &
	$\mathrm{dB.s}^{- 2}\text{Hz}$\tabularnewline
	\bottomrule
\end{longtable}
\end{table}

\subsection{Time-series estimates}
\label{ss:real-time}

\begin{figure}[h]
	\centering
	\includegraphics[width=0.47\linewidth]{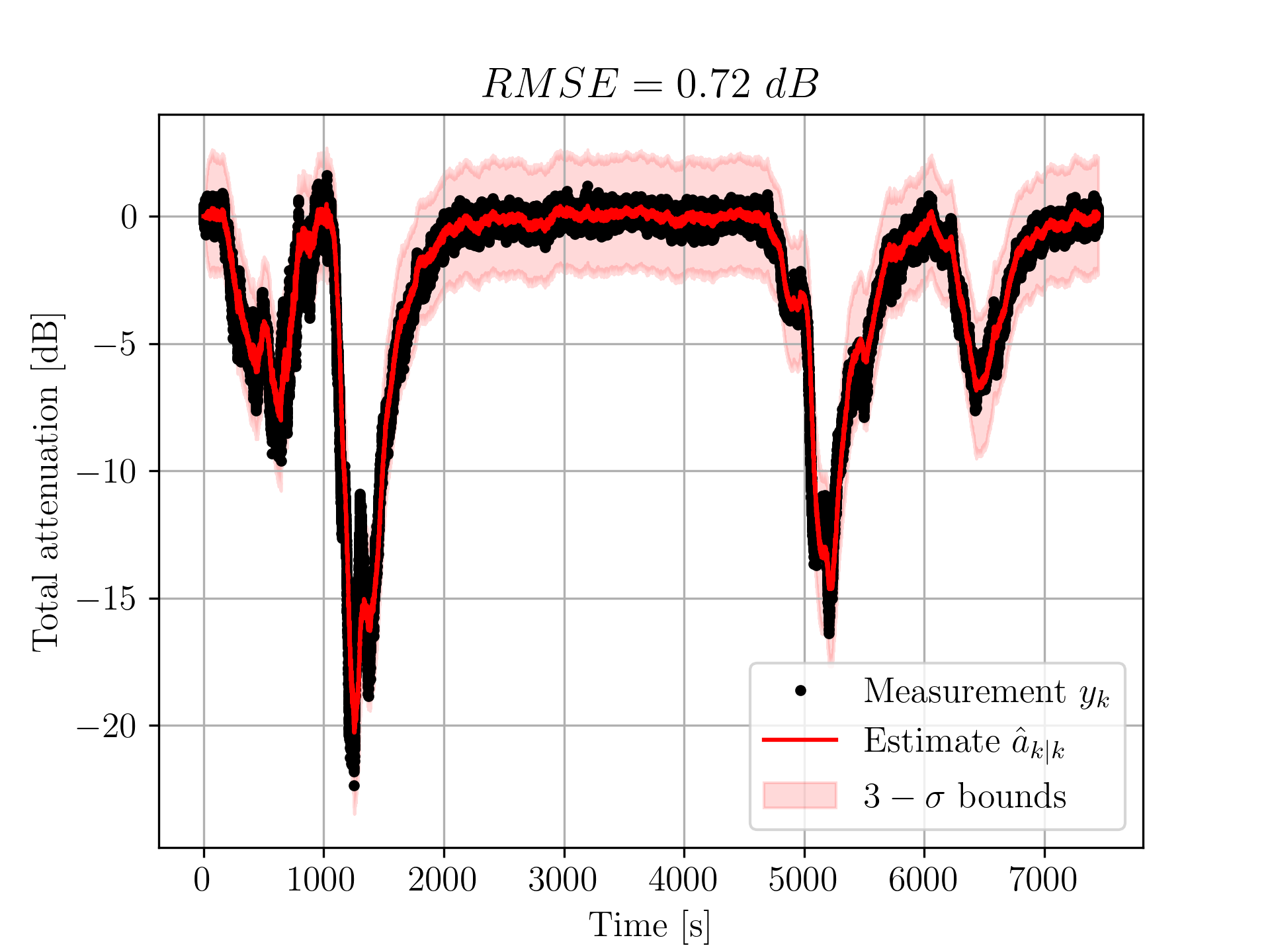}
	\caption{Attenuation measurement and estimates.}
	\label{fig:ts_att}
\end{figure}

Here, we present a practical use-case of SciFi over a tropospheric event in order to show its filtering capabilities.
We plotted in Figure \ref{fig:ts_att} the overlap of the attenuation estimates
${\widehat{a}}_{k}$ and the measurements $y_{k}$ during the event. 
We remark that our filter is smoothing efficiently the
measurements and remove outliers due to scintillation. 
On the other
hand, the shape of the event is preserved by SciFi, which also provides
confidence intervals along the time series. These intervals are shaded
in red in Figure \ref{fig:ts_att} and computed using the $3\sigma$ bound method.

Namely, for a given state $s \in \{a,\dot{a}\}$ these bounds are
given by 
\begin{equation}
	b_{s, k, \pm} = \hat{s}_{k|k} + 3\sqrt{P_{\text{ss}}},
	\label{eq:bounds}
\end{equation}
where $P_{\text{ss}}$
is the variance of the estimate $\hat{s}_{k|k}$ provided by the
SciFi algorithm. For instance, if $s=0$ the $(1,1)$ coefficient of the matrix
$\mathbf{P}_{k|k}$ defined in \eqref{eq:col_propag} would be selected.
This bounding
represents a 99.7\% estimated probability of inclusion, \textit{i.e.,} $\mathbb{P}(s_k \in [b_{s, k, -},b_{s, k, +}])=0.997$,
 assuming the estimate $\hat{s}_{k|k}$ being Gaussian centered.

\begin{figure}[b]
	\centering
	\includegraphics[width=0.47\linewidth]{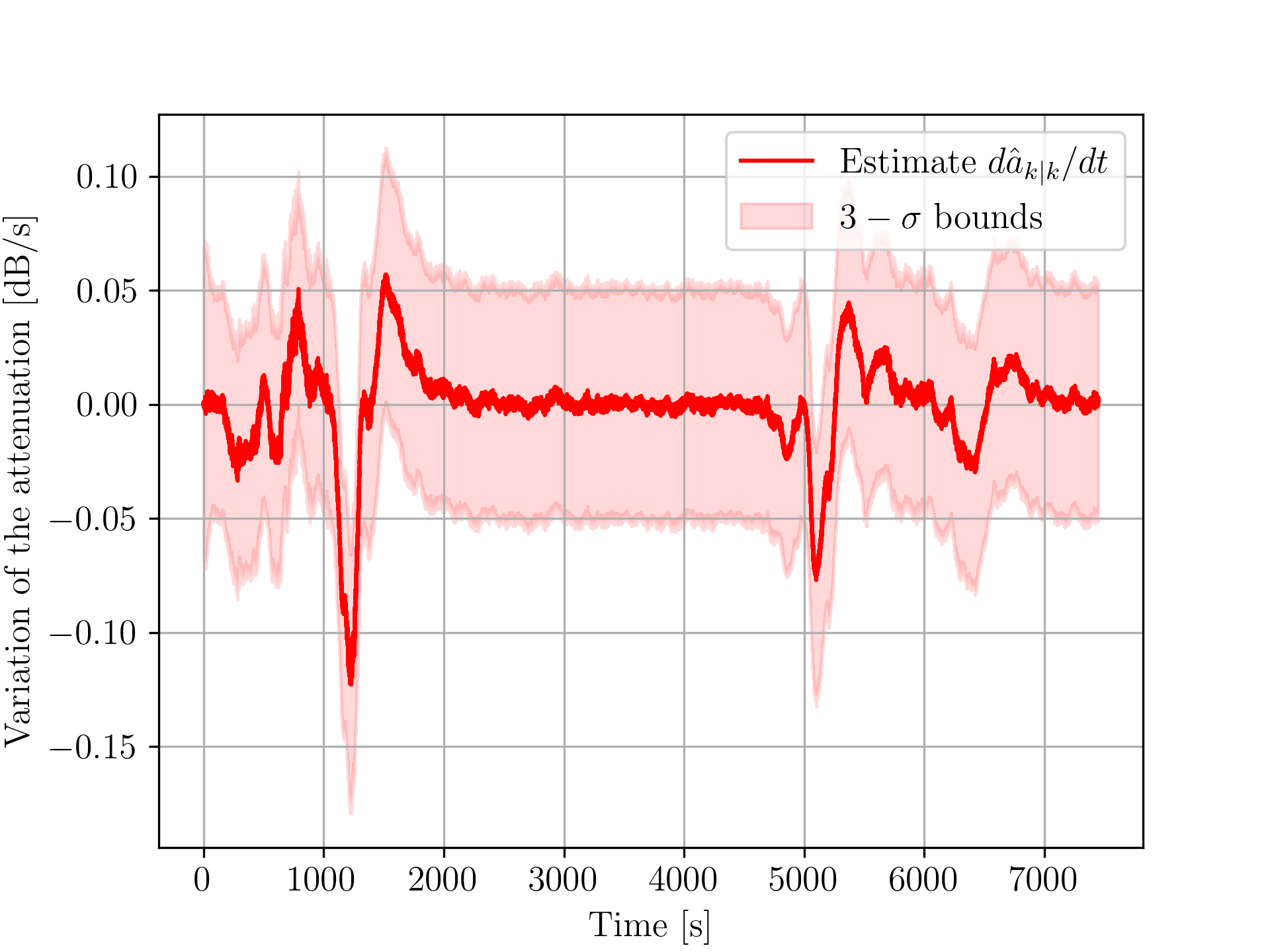}
	\caption{Attenuation slope estimates.}	
	\label{fig:ts_slope}
\end{figure}

We remark that the interval formed by these bounds in Figure \ref{fig:ts_att} is larger when the mean of the
measurements becomes highly variable, \textit{i.e.}, when the attenuation slope is high. This fact is explained by the simplicity of the model \eqref{eq:continuous_time_attenuation_dynamics}, which is a single integrator. Indeed, the uncertainties are introduced by both the process noise (considered as driving the slope derivative $\ddot{a}$) and high values of the innovation that occurs when the slope is significantly varying.
 One way to reduce these confidence bounds would be to include extrinsic information (hydrometric data, weather data) in the attenuation dynamic model, which is outside the scope of our work. On the other hand, these bounds tend to be conservative when the attenuation is close to zero owing of the aggressive tuning of the process noise PSD $\sigma_{\text{ww}}$. This uncertainty data can be exploited in order to give safety margins for FMT triggering.

On the other hand, Figure \ref{fig:ts_slope} shows the results of the estimation of the
attenuation slope $\dot{a}$ with a superposition of its
$3\sigma$ bound defined by \eqref{eq:bounds} considering $s:=\dot{a}$. Note that the
estimate of the slope is noisier than that of the attenuation, since
this variable is not directly observed and that no extrinsic data is
provided in the model to predict its trend. Moreover, the Brownian motion model  \eqref{eq:continuous_time_attenuation_dynamics} directly introduces noise sensitivity in the state estimate $\hat{\dot{a}}$. 

Overall, this application of SciFi yielded satisfactory results on real data, using few matrix operations that can be performed in real time thanks to moderate computational cost. 

\subsection{Statistical results}
\label{ss:statistic}

Here, we present statistical results of attenuation and slope estimates obtained from measurements with the same setting in Toulouse, as presented in Table \ref{tab:input}. The results have been processed using five years of Q/V beacon data between 2018 and 2022. We have compared the results of Scifi (abbreviated by KF) and the standard Low-Pass method (resp. LP) on a statistical basis. The latter computes $\hat{a}_\text{LP}$ using a fifth order Butterworth filter with a cutoff frequency of $2.5\times 10^{-2}~\mathrm{Hz}$ as in the paper of Comisso \textit{et al.} \cite{comisso_tropospheric_2023}. The LP slope estimate is given by a straightforward simple differentiation scheme 
\begin{equation}
	\hat{\dot{a}}_\text{k,LP}=(a_{\text{LP},k}-a_{\text{LP},k-1})f_s,
	\label{eq:diff}
\end{equation}
where $f_s=10~\mathrm{Hz}$ is the sampling frequency.

\begin{figure}[h]
	\centering
	\includegraphics[width=0.47\linewidth]{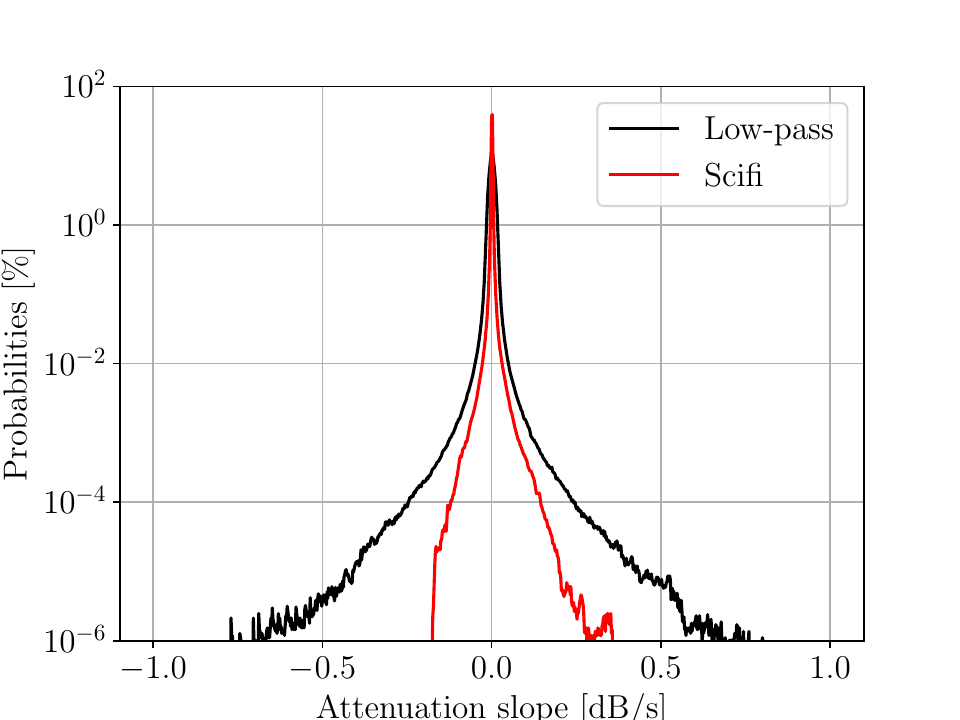}
	\caption{PDF of the attenuation slope using Scifi (KF) and the Low-pass method (LP) for the data specified in Table \ref{tab:input}.}
	\label{fig:dotpdf}
\end{figure}

Figure \ref{fig:dotpdf} shows the empirical attenuation slope Probability Density Function (PDF) computed with the two methods. We represented in black the PDF of $\hat{\dot{a}}_\textbf{LP}$ (baseline) obtained using \eqref{eq:diff} after LP filtering and in red the result of SciFi $\hat{\dot{a}}$ leveraging the steps of Algorithm \ref{alg:scifi}. This color code is used throughout the rest of this section. First, we can remark that the LP seems to introduce slope overestimation. 
Indeed, we can see strong fade slope magnitude above $0.5~\mathrm{dB/s}$, though at a low probability, which seems unlikely to happen in temperate climate. 
In fact, the finite difference scheme tend to amplify high frequency components that can occur despite the application of the Butterworth filter. Increasing the order of the filter removes these artifacts but introduces an additional lag in the estimation process. 
On the other hand, SciFi tends to output a sharper slope PDF, mitigating these numerical artifacts. Indeed, the KF framework estimates the states PDF (assumed Gaussians) and in particular its expectation over time, which prevents abrupt changes in the time series. 

\pagebreak
\begin{figure}[t]
	\centering
	\includegraphics[width=0.47\linewidth]{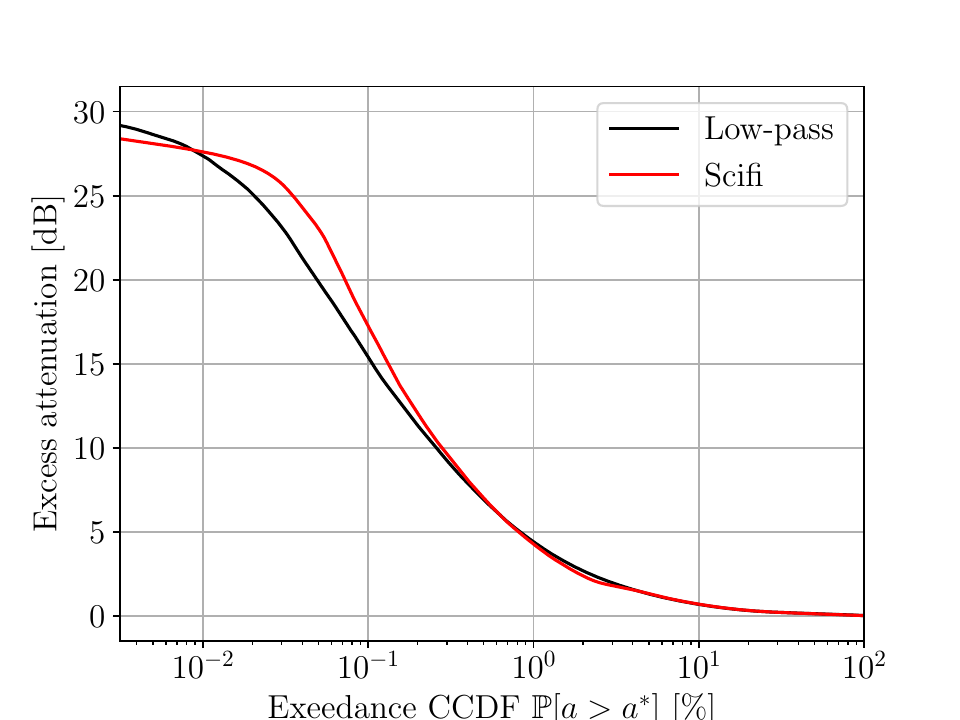}
	\caption{Attenuation CCDFs obtained with LP and Scifi methods.}
	\label{fig:attccdf}
\end{figure}

Figure \ref{fig:attccdf} presents the CCDF for the attenuation estimates $\hat{a}$. As expected, the events having more than $1\%$ of probability are similarly detected. However events between $10^{-1}\%$ and $10^{-2}\%$ are underestimated by the LP approach with respect to SciFi. This fact comes from the abrupt averaging of the time series when using a high order Butterworth filter, while the KF tends to track events more accurately when the slope is moderate. 

\begin{figure}[h]
	\centering
	\includegraphics[width=0.47\linewidth]{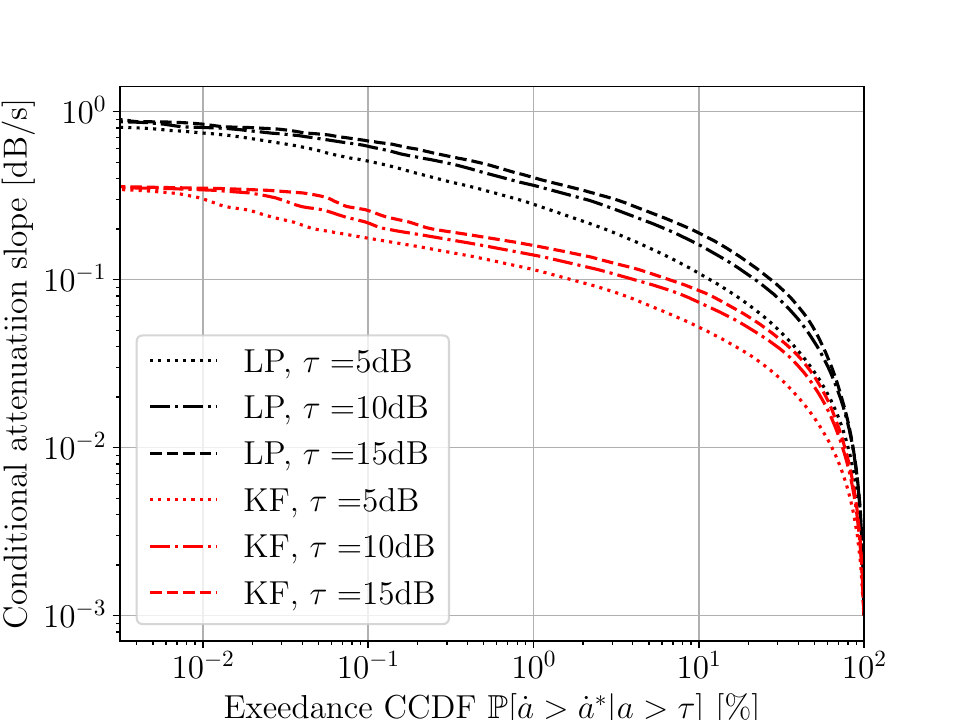}
	\caption{Conditional CCDF of the attenuation slope, obtained with LP and Scifi methods, with respect to the attenuation threshold $\tau$ defined in \eqref{eq:cond_ccdf}.}
	\label{fig:conddotccdf}
\end{figure}

Finally, the conditional statistics of the absolute value of the slope as in \itu{1623-1} \cite{itu-r_recommendation_2005} are presented in Figure \ref{fig:conddotccdf}. The definition of this CCDF, introduced in the work of Van de Kamp \cite{van_de_kamp_statistical_2003} is defined as follows
\begin{equation}
	 \mathbb{P}[|\dot{a}|>\dot{a}^*|a>\tau] = \mathbb{P}[\dot{a}>\dot{a}^* \cap a>\tau] (\mathbb{P}[a>\tau])^{-1}
\label{eq:cond_ccdf}
\end{equation}
where $\dot{a}^*$ is presented in ordinate and $\mathbb{P}[|\dot{a}|>\dot{a}^*|a>\tau]$ in abscissa of Figure \ref{fig:conddotccdf}. The variable $\tau$ is a conditional threshold that prevents incorporating small attenuation values in the CCDF computation, which have least interest for FMT techniques. One can also note that this condition prevents differentiation noises when using LP techniques, \textit{i.e.,} small values of excess attenuation are essentially a contribution of the tropospheric scintillation. Thus, these components, despite being filtered, are then amplified while computing the slope with \eqref{eq:diff}. Here again, we remark that the LP baseline tend to overestimate the attenuation slope with respect to SciFi due to its more aggressive differentiation method. Also, we can question the reliability of LP estimation, even with such conditional triggering, that here again yields slopes over $0.5~\mathrm{dB/s}$.

In this Subsection, we shown that SciFi is also able to extract data statistics over a large period of time and, again, with a moderate computational cost. Therefore, SciFi is an interesting tool to compute statistical profiles of attenuation without scintillation. Moreover, one can note that the scintillation distribution can also be extracted thanks to SciFi computing the difference $y-\hat{a}$.

\FloatBarrier 
\section{Conclusion and perspectives}
\label{conclusion-and-perspectives}

In summary, the SciFi algorithm provides an innovative approach to remove scintillation from attenuation data using a stochastic linear observer framework instead of transform-based filters. This algorithm is fed by a simple attenuation model and a fine stochastic modeling of the noise based on ITU-R Recommendations, which allows a reliable estimation of the tropospheric excess attenuation $a$. Moreover, SciFi also estimates the slope of the filtered attenuation $\dot{a}$ while providing simultaneously the uncertainties, by computing the covariance of its tracked states $a$ and $\dot{a}$. Therefore, this information could be used with benefit to design more robust Fade Mitigation Techniques policies for VHTS systems.

Thanks to the moderate computational cost of SciFi, it can be used to perform short-term
predictions, since the Kalman prediction step can be computed at each time point. Finally, to the best of our knowledge, this approach of simultaneously
performing scintillation removal and attenuation slope estimation, is novel. On the other hand, our algorithm is a promising tool to compute tropospheric attenuation statistics since it requires very few parameterization steps to implement.  

Future work will include an implementation of SciFi in a large-scale satellite communication system simulator, using it as an input for site diversity decision and control processes. Indeed, both the uncertainties and the trend given by the slope estimation are of high value to design novel switching policies that aim to minimize the number of switches while increasing the system availability.

\bmsection*{Acknowledgments}
The authors would like to thank Dr. Laurent Feral, head of the Propagation, Environment and Radiocommunications team of ONERA, for his careful proofreading of the extended version of the manuscript and Alkan Yerebasmaz M.S. student at Université Pierre-et-Marie-Curie, France, for his advice on how to improve the readability of SciFi during his internship at ONERA.

\bibliography{biblio.bib}


\bmsection*{Author Biographies}

\newcommand{\pict}[1]{\includegraphics[height=76pt]{fig/staff/#1.jpg}}

\begin{biography}{\pict{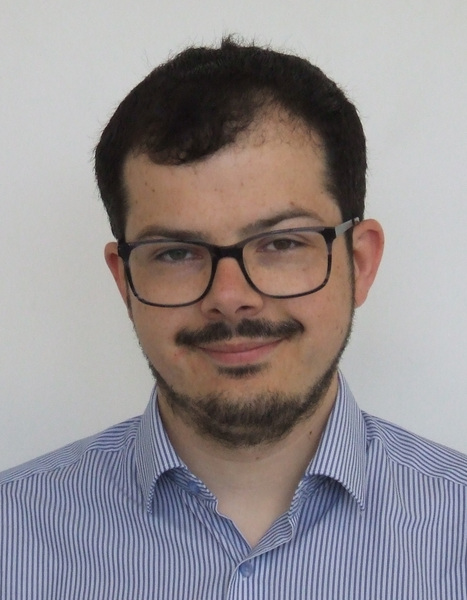}}{
{\textbf{Justin Cano} was born in 1994 in Nîmes, France. He graduated in 2019 with dual degrees: a Master of Engineering from Ecole Centrale Mediterannée, Marseille, France, and a Master of Science in Electrical Engineering from Polytechnique Montréal, Quebec, Canada. In 2023, he earned a \textit{Ph.D.} in Electrical Engineering, awarded \textit{cum laude} from both ISAE-Supaéro (France) and Polytechnique Montréal in 2023. Dr. Cano is now with the Electromagnetism and Radar Department (DEMR) of ONERA, France, where he specializes in Satellite Communication systems, with a focus on developing new Fade Mitigation Techniques.}}
\end{biography}

\begin{biography}{\pict{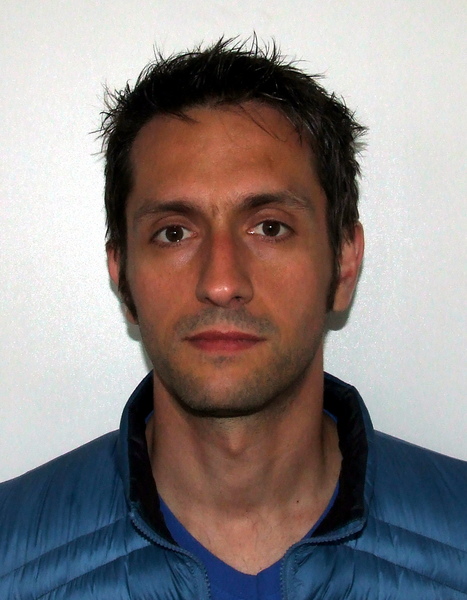}}{
		{\textbf{Julien Queyrel} was born in 1983 in Gap, France. He graduated in 2007 with a Master of Engineering from École nationale Supérieure de l’Aéronautique et de l’Espace (ISAE) in Toulouse, France, and a Research Master’s in Astrophysics from Université de Toulouse. In 2010, he earned a Ph.D. in Astrophysics from Université de Toulouse. Following his doctoral studies, he worked as a subcontractor for the French National Space Agency (CNES) for seven years, contributing to astrophysics space missions and ELINT satellite projects. In 2018, he joined ONERA, specializing in tropospheric microwave propagation with an emphasis on stochastic methods and high-resolution meteorological models.}}
\end{biography}

\begin{biography}{\pict{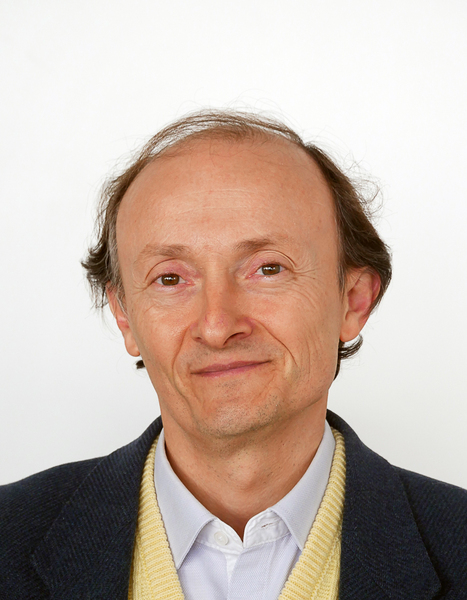}}{
		{\textbf{Laurent Castanet}  was born in Dunkerque, France in 1967. He received the B.S. Degree in Microwave Engineering from Telecom Bretagne in 1991, the M.S. Degree in Space Communication Systems from Telecom Paris in 1992 and the PhD Degree from SUPAERO on “Fade Mitigation Techniques for new SatCom systems operating at Ka and V bands”.
			
			Since 1994, he has been working as research engineer at ONERA Toulouse on radiowave propagation in particular for high frequency satellite communication systems and on the impact of the environment on system performances using Fade Mitigation or signal processing Techniques.
			
			He is the Head of the French delegation to ITU-R Study Group 3 and vice-chair of WP3J. He teaches radiowave propagation and radio link analysis in French Engineering schools, at University of Toulouse and at EuroSAE training institution.
			}}
\end{biography}

\begin{biography}{\pict{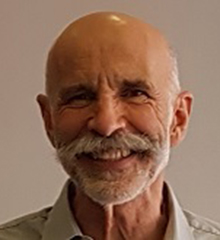}}{
		{\textbf{Michel Bousquet.} Prof. retired from ISAE-SUPAERO, the French Aerospace Engineering Institute of Higher Education where he directed master programmes and was leading a research team in communications and navigation. Born in 1951, holding degrees from University of Toulouse and Supaero, he is still active in non-profit organizations involved in research activities or support to students. With over forty years of teaching and research experience, participation to many European and ESA R\&D projects, he is the author/co-author of many papers and textbooks. Standing on the boards of European research programmes (COST, SatNEx Network of Excellence...). visiting lecturer with several foreign universities, organizer of numerous seminars on satellite systems, he is recipient of the 2019 AIAA Aerospace Communication award for his ``For his outstanding contribution and promotion of education and proliferation of knowledge on aerospace communication and navigation''.}}
\end{biography}

\end{document}